\documentclass[%
 reprint,
superscriptaddress,
 amsmath,amssymb,
prb,
floatfix,
]{revtex4-1}

\usepackage{siunitx}
\usepackage{graphicx}
\usepackage{dcolumn}
\usepackage{bm}
\usepackage{textcomp}
\usepackage{gensymb}

\usepackage{multirow}
\usepackage{amsmath,amssymb}
\usepackage{soul}
\usepackage{hyperref}
\usepackage{xcolor}
\usepackage{verbatim}
\hypersetup{
    colorlinks=true,
    linkcolor=blue,
    citecolor=blue,
    urlcolor=blue,
}

\begin{document}

\preprint{APS/123-QED}

\title[manuscript in preparation]{Mapping out the spin-wave modes of constriction-based spin Hall nano-oscillators in weak in-plane fields}

\author{Hamid~Mazraati}
\affiliation{NanOsc AB, Kista 164 40, Sweden}
\affiliation{Department of Applied Physics, School of Engineering Sciences, KTH Royal Institute of Technology, Electrum 229, SE-16440 Kista, Sweden}

\author{Seyyed~Ruhollah~Etesami}
\affiliation{Department of Physics, University of Gothenburg, 412 96, Gothenburg, Sweden}

\author{Seyed~Amir~Hossein~Banuazizi}
\affiliation{Department of Applied Physics, School of Engineering Sciences, KTH Royal Institute of Technology, Electrum 229, SE-16440 Kista, Sweden}

\author{Sunjae~Chung}
\affiliation{Department of Applied Physics, School of Engineering Sciences, KTH Royal Institute of Technology, Electrum 229, SE-16440 Kista, Sweden}
\affiliation{Department of Physics, University of Gothenburg, 412 96, Gothenburg, Sweden}

\author{Afshin~Houshang}
\affiliation{NanOsc AB, Kista 164 40, Sweden}
\affiliation{Department of Physics, University of Gothenburg, 412 96, Gothenburg, Sweden}

\author{Ahmad~A.~Awad}
\affiliation{NanOsc AB, Kista 164 40, Sweden}
\affiliation{Department of Physics, University of Gothenburg, 412 96, Gothenburg, Sweden}

\author{Mykola~Dvornik}
\affiliation{NanOsc AB, Kista 164 40, Sweden}
\affiliation{Department of Physics, University of Gothenburg, 412 96, Gothenburg, Sweden}

\author{Johan~\AA{}kerman}
\affiliation{NanOsc AB, Kista 164 40, Sweden}
\affiliation{Department of Applied Physics, School of Engineering Sciences, KTH Royal Institute of Technology, Electrum 229, SE-16440 Kista, Sweden}
\affiliation{Department of Physics, University of Gothenburg, 412 96, Gothenburg, Sweden}

\date{\today}

\begin{abstract}

We experimentally study the auto-oscillating spin-wave modes in NiFe/$\beta-$W constriction-based spin Hall nano-oscillators as a function of bias current, in-plane applied field strength, and azimuthal field angle, in the low-field range of 40--80 mT. 
We observe two different spin-wave modes: \emph{i}) a linear-like mode confined to the minima of the internal field near the edges of the nanoconstriction, with weak frequency dependencies on the bias current and the applied field angle, and \emph{ii}) a second, lower frequency mode that has significantly higher threshold current and stronger frequency dependencies on both bias current and the external field angle. Our micromagnetic modeling qualitatively reproduces the experimental data and reveals that the second mode is a spin-wave bullet and that the SHNO mode hops between the two modes, resulting in a substantial increase in linewidths. In contrast to the linear-like mode, the bullet is localized in the middle of the constriction and shrinks with increasing bias current. Utilizing intrinsic frequency doubling at zero field angle we can reach frequencies above 9 GHz in fields as low as 40 mT, which is important for the development of low-field spintronic oscillators with applications in microwave signal generation and neuromorphic computing.

\begin{description}
\item[PACS numbers]
\end{description}
\end{abstract}

\pacs{Valid PACS appear here}
\maketitle


\section{\label{sec:Intro}Introduction}

Spin torque nano-oscillators---microwave signal generating devices based on spin-wave auto-oscillations---are of great interest for many kinds of nanoscale applications as they provide highly coherent and widely tunable microwave signals at room temperature. \cite{chen2016ieeerev} Recently, they have been succeeded by so-called spin Hall nano-oscillators (SHNOs), which utilize the spin Hall effect\cite{Hirsch1999,Zhang2000,Kato2004,Wunderlich2005,Saitoh2006, Valenzuela2006a} to generate microwave signals of similar quality.\cite{Demidov2012b} To date, a variety of SHNO geometries and material compositions have been proposed.\cite{Demidov2012b,Liu2013,Zholud2014,Duan2014b, Collet2016,Ranjbar2014,C6NR07903B,Pai2012,doi:10.1063/1.4907240,Mazraati2016,doi:10.1063/1.5022049,Spicer2018,Yin2018} Most recently, a constriction-based SHNO was developed with the particular advantages of having a rather simple fabrication process and relatively low driving current.\cite{Demidov2014} Later, the mutual synchronization of multiple constriction-based SHNOs was experimentally demonstrated for strong oblique magnetic fields, and substantial improvements in the output power and quality factor were observed.\cite{Awad2016} Thanks to the unprecedented ability of constriction SHNOs to phase-lock with each other, they may also be utilized for future spintronic neuromorphic computing devices.\cite{Torrejon2017Nature,Dvornik2017} However, most practical applications require these devices to operate in either zero or weak applied magnetic fields. A deeper understanding of the SHNO dynamics in such regimes is thus necessary for further developments.

Demidov \textit{et al.}\cite{Demidov2014}~demonstrated that, for an in-plane field of 40~mT, applied at \ang{40}~\emph{w.r.t} the drive current, constriction-based SHNOs exhibit a single auto-oscillating mode over a wide range of applied currents with weak negative frequency versus current tunability. Dvornik \textit{et al.}\cite{Dvornik2018prappl} later showed that such auto-oscillations emerge from the linear localized mode of the nanoconstriction. Although a transition to multimode operation with substantial line broadening had also been observed in Ref.~\onlinecite{Demidov2014}, neither was it discussed in detail nor was the origin of the additional peaks explained. Finally, optimization of the in-plane field angle is essential to achieve high output power and robust mutual synchronization of these devices in in-plane fields.\cite{Kendziorczyk2016prb}

In this work, we report for the first time on angular-resolved measurements of constriction based SHNO microwave signal generation under weak in-plane fields, $H_{\mathrm{IP}}$. We observe both a linear-like mode confined to the minima of the internal field near the edges of the nanoconstriction, and a lower frequency spin wave bullet mode localized in the middle of the constriction. Our micromagnetic simulations suggest that the SHNO hops rapidly between these two modes, consistent with the much larger linewidths observed in this regime. Finally, we use intrinsic frequency doubling to achieve frequencies exceeding 9 GHz in fields as low as 40 mT.

\section{\label{sec:Exp}Experiment}
\subsection{\label{sec:DevFabrication}Device fabrication and measurement setup}

The SHNO stack, consisting of A NiFe(5nm)/$\beta$-W(5nm) bilayer, was prepared on a c-plane sapphire substrate using dc/rf magnetron sputtering in a 2.5 mTorr Argon atmosphere, in an ultra-high vacuum chamber (base pressure below $1\times 10^{-8}$ mTorr). It was then patterned into an array of $4~\mu$m $\times$ $12~\mu$m rectangular mesas using photolithography and Argon ion milling. Nanoconstrictions with a width of 150~nm were subsequently fabricated in the center of these mesas by a combination of electron-beam lithography and dry etching. To determine the magnetic characteristics of the stack using spin-torque-induced ferromagnetic resonance (ST-FMR) measurements, 6~$\mu$m-wide bars were simultaneously fabricated next to the SHNOs. Finally, a conventional ground--signal--ground (GSG) waveguide and electrical contact pads for broad frequency range microwave measurement were fabricated by lift-off photolithography and Cu/Au sputtering on top of both the nanoconstrictions and the bars.

\begin{figure}[b]
    \centering
\includegraphics[trim=0cm 0.5cm 0cm 0cm, clip=true,width=3.4in]{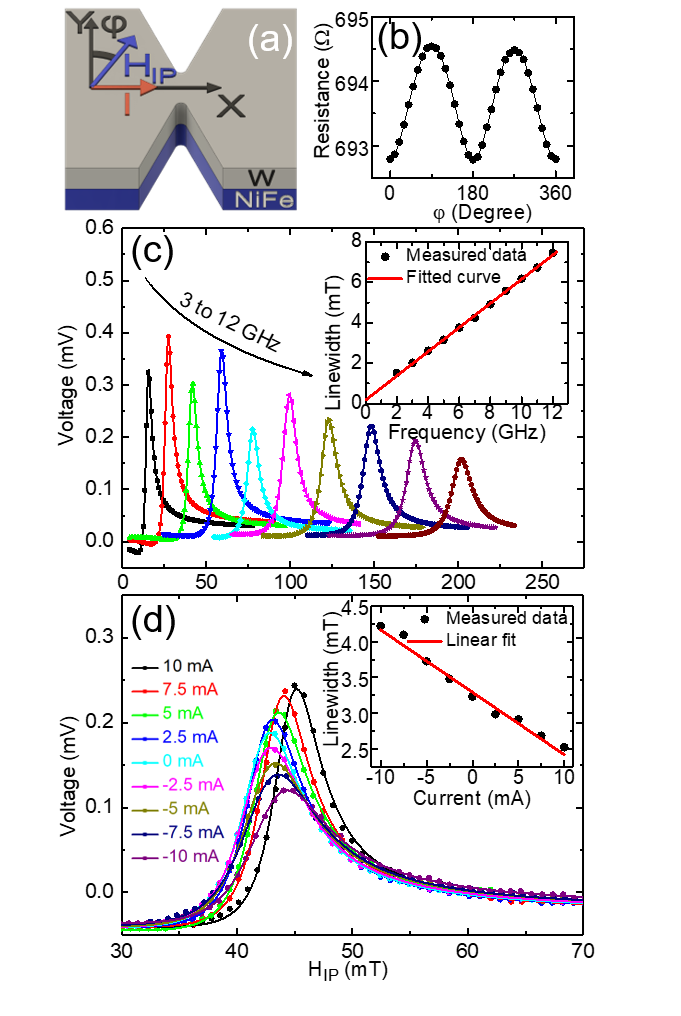}
\caption{\label{fig1} (Color online) (a) Schematic of the nanoconstriction-based SHNO and the configuration of the in-plane field and the current, (b) the angular dependent magnetoresistance of the stack, (c) ST-FMR peaks at $I_{\mathrm{dc}}=0$~mA for different microwave frequencies, and (d) ST-FMR peaks for $f=5$~GHz at different dc current. Solid lines show Lorentzian fits. The insets show the extracted linewidths of ST-FMR peaks and their linear fits.}
\end{figure}

Fig.~\ref{fig1}~(a) shows a schematic of the device structure, including the directions of the applied in-plane field and current: the field angles $\varphi=0\degree$ and $+90\degree$ are along the +$y$ and +$x$ axes, respectively. A negative (positive) current represents electrons flowing along the $+(-)x$ direction. 

The magnetoresistance of the SHNO vs.~ in-plane field angle is shown in Fig.~\ref{fig1}~(b) revealing an AMR ratio of 0.26 $\%$, similar to literature values for thin NiFi films.\cite{1058782} We carried out ST-FMR measurement on bars using the homodyne detection approach.\cite{Sankey2006a,Sankey2007b,Chen2009,Cheng2013,Collet2016,Fazlali2016prb} A 313 Hz--pulse--modulated microwave signal was applied alongside a direct current through a bias-tee, and the modulated signal was then detected through the same bias-tee and analyzed using a lock-in amplifier. The applied field was swept from 250~mT to 0~mT while the frequency of the input microwave signal and the level of the direct current were fixed.

Microwave measurements were carried out in a custom-built setup. While a direct current was injected through the constriction area of the SHNO under an in-plane field, the auto-oscillation microwave signal was acquired by a spectrum analyzer after being amplified 35~dB using a broadband low-noise amplifier. All measurements were performed at room temperature. 

\subsection{\label{sec:Results}Results}

Fig.~\ref{fig1}~(c) shows the ST-FMR spectra at different microwave frequencies from 3 to 12~GHz (dots), with each spectrum well fitted to a sum of one symmetric and one asymmetric Lorentzian (solid lines).\cite{Liu2011a} The extracted frequencies of the resonance peaks fit well to the Kittel formula~\cite{Kittel1948}, leading to an effective magnetization of $\mu_{\mathrm{0}}M_{\mathrm{eff}}=0.71$~T and a gyromagnetic ratio of $\gamma/2\pi=28$~GHz/T. The inset to Fig.~\ref{fig1}~(c) shows the extracted linewidths of the corresponding peaks (black dots) and their fit to the linear model (solid red line). The obtained value of the Gilbert damping is $\alpha=0.016$.

Fig.~\ref{fig1}~(d) shows the ST-FMR spectra for a range of bias currents from +10~mA to -10~mA, measured at a fixed microwave frequency of 5~GHz. The extracted linewidth vs.~current behaviour and a fit to the linear model\cite{Demasius2016,Ando2008c,Liu2011a} are shown in the inset. The extracted value of the spin Hall efficiency, defined as the ratio of the absorbed spin and the charge density currents, is $\xi_{\mathrm{SH}}=-0.385$, which is significantly higher than in the case of NiFe/Pt stacks.\cite{Demidov2012b,Liu2013, Ulrichs2013,Demidov2014,Duan2014b,Ranjbar2014,Zholud2014,Collet2016}

\begin{figure}[t]
    \centering
\includegraphics[trim=0cm 0.5cm 0cm 0cm, clip=true,width=3.4in]{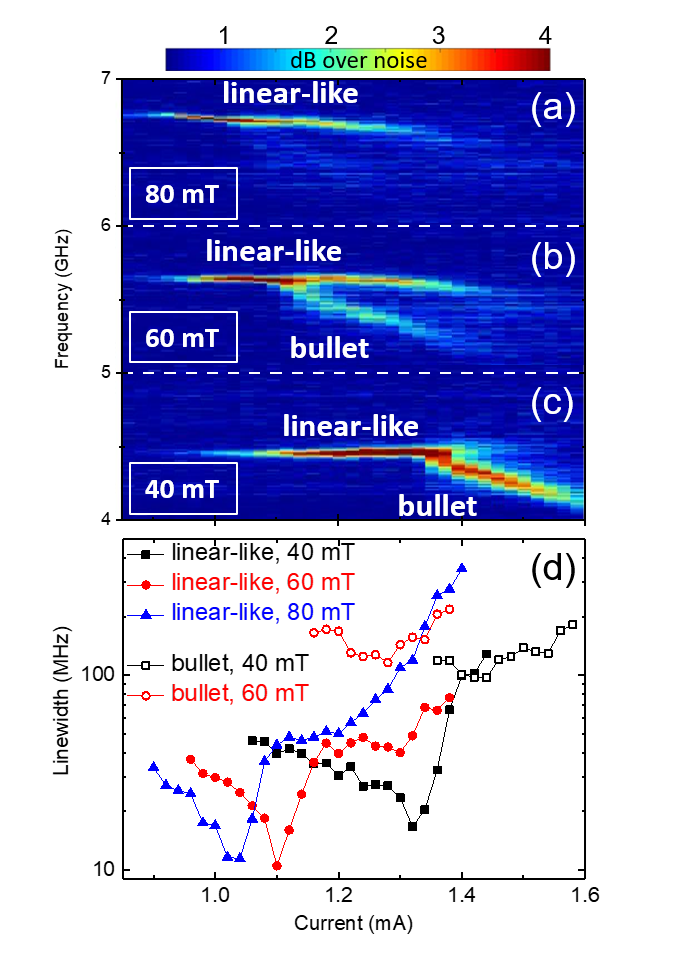}
\caption{\label{fig2} (Color online) Current dependent auto-oscillation power spectral density for the field of (a) $H_{\mathrm{IP}}=0.04$, (b) 0.06, and (c) 0.08~T applied along $\varphi=30\degree$. (d) Fitted linewidth of the corresponding resonant peaks.}
\end{figure}

The power spectral densities (PSDs) of the SHNO vs.~bias current for the fields of 40~mT, 60~mT, and 80~mT applied $\varphi=30\degree$ in-plane are shown in Fig.~\ref{fig2}~(a)--(c). For $\mu_0H_{\mathrm{IP}}=80$~mT, there is a single dominant spin-wave mode that, according to Ref.~\onlinecite{Dvornik2018prappl}, should originate from the linear magnonic edge mode of the constriction (hence the label ``linear-like mode''). In contrast to the uniform ferromagnetic resonance of in-plane magnetized films, where the frequency decreases with the amplitude of precession, the observed auto-oscillations experience a nonmonotonic frequency vs.~current behaviour. At lower currents, the frequency of the mode is almost constant while its linewidth decreases with increasing current. However, at a certain field-dependent current the mode shows a redshift (negative nonlinearity), and the linewidth starts to increase. At the same time, traces of a lower frequency and larger linewidth signal can be seen in Fig.~\ref{fig2}(a). These are the signatures of the so-called spin-wave bullet---a nonlinear and nontopological self-localized mode nucleated in regions of negative non-linearity.\cite{PhysRevLett.95.237201,Demidov2012b,PhysRevLett.105.217204,Boneti2012,Spicer2018b,Jungfleisch2016}. They become more apparent when the applied field is reduced to 60~mT (Fig.~\ref{fig2}(b)) and eventually dominate at 40~mT (Fig.~\ref{fig2}(c)). Our experimental data suggest that the contribution of the negative nonlinearity could increase with the applied field, shifting its onset current downwards from approximately 1.4~mA at 40~mT to 1~mA at 80~mT.

Generally, the linear-like and the bullet modes cannot coexist when overlapping in space, so the presence of both signals in the measured spectra is likely due to mode hopping.\cite{Boneti2012,MuduliPrl2012,Muduli2012prb, Iacocca2014prb,Heinonen2013,Sharma2014} The reduced stability and broader linewidth of the bullet mode at higher fields indicate an increase in the nonlinearity and the suppression of the nonlinear magnetic losses\cite{Tiberkevich2007prb} that limit the auto-oscillation amplitude.

\begin{figure}[t]
    \centering
\includegraphics[width=3.4in]{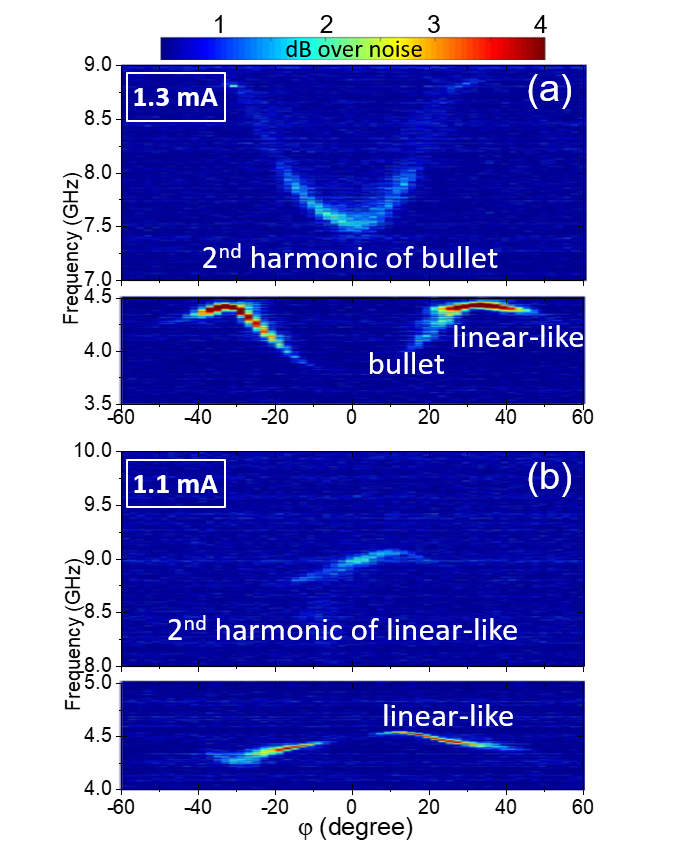}
\caption{\label{fig3} (Color online) PSD map of the fundamental and second harmonic of the modes versus the in-plane angle of the field for $\mu_0\mathrm{H_{IP}}=40$~mT at (a) $I_{\mathrm{dc}}=1.3$~mA, and (b) $I_{\mathrm{dc}}=1.1$~mA }
\end{figure}

The fundamental and second harmonics of both modes as a function of the in-plane angle of the applied field of constant strength, $H_{\mathrm{IP}}=40$~mT, are plotted in Fig.~\ref{fig3}~(a) and (b) at $I_{\mathrm{dc}}=1.3$~mA and 1.1~mA, respectively. The slight asymmetry of the responses with respect to the field angle is likely due to fabrication-related shape imperfections or the slight misalignment of the sample with respect to the center of the magnet. We observed no auto-oscillations for angles beyond $\lvert \varphi \rvert=45 \degree$, either because the edge modes (a) disappear due to the suppression of the spin-wave wells or (b) do not get excited as they move away from the constriction, experiencing less spin-current density. The former is unlikely since it would be accompanied by a considerable increase in frequency which we do not see experimentally. In fact, the frequency of the linear-like mode depends only weakly on the in-plane angle of the applied field, suggesting minimal changes to its localization depth.

The bullet mode was observed at higher currents (Fig.~\ref{fig3}~(a)), and disappeared as the current dropped (Fig.~\ref{fig3}~(b)), similar to the PSD maps shown in Fig.~\ref{fig2}. In contrast to the linear-like mode, the frequency of the bullet mode depends strongly on the field angle and decreases with the angle of the external field. This may be attributed to the angular dependence of the damping-like torque. At low angles, the magnetization vector points mostly antiparallel to the polarization of the spin current resulting in higher spin torque efficiency, larger bullet amplitude, and thus a higher nonlinear frequency redshift.

It is worth noting that the fundamental harmonic for any of the modes were not detected for small field angles. This is a consequence of the first derivative of the AMR curve (Fig.~\ref{fig1}b) approaching zero at $\varphi=0\degree$. However, both modes were instead clearly observed by their signals at twice their original frequencies, since the second derivative of the AMR curve has a maximum in the vicinity of $\varphi=0\degree$.\cite{Muduli2011jap} As a consequence, using this intrinsic frequency doubling, we can reach very high frequencies already at very low fields.

\section{\label{sec:Simulation}Micromagnetic simulations}

In order to investigate the physics behind the experimentally observed spin-wave modes, we carried out micromagnetic simulations using MuMax3\cite{Vansteenkiste2014}. The material parameters used in the simulations---such as the NiFe saturation magnetization, the Gilbert damping, the gyromagnetic ratio, and the spin Hall efficiency of the bilayer---were obtained directly from the ST-FMR measurements. An exchange stiffness of $A_{\mathrm{ex}}=10^{-12}$~J/m was considered for NiFe. We ran the simulations for a geometry with a lateral size of 2000$\times$2000~nm$^2$ (large enough to avoid boundary effects) and a thickness of 5~nm, which is similar to the thickness of the ferromagnetic layer in the real sample. The distributions of the direct charge current and the corresponding Oersted field were obtained using COMSOL Multiphysics$\textsuperscript{\textregistered}$ software \cite{Dvornik2018prappl} for a NiFe/W bilayer with resistivity values of 0.90~$\mu \Omega \cdot$m and 2.12~$\mu \Omega \cdot$m for NiFe and W, respectively. The auto-oscillation spectra were obtained by applying a Fast Fourier Transform (FFT) to the net $y$ component of the total magnetization, simulated over 1000~ns. We assumed that the spin-current polarization equals $P=1$ and is independent of the angle between the directions of magnetization and of spin-current polarization.

\begin{figure}[t]
    \centering
\includegraphics[trim=0cm 0.5cm 0cm 0cm, clip=true,width=3.4in]{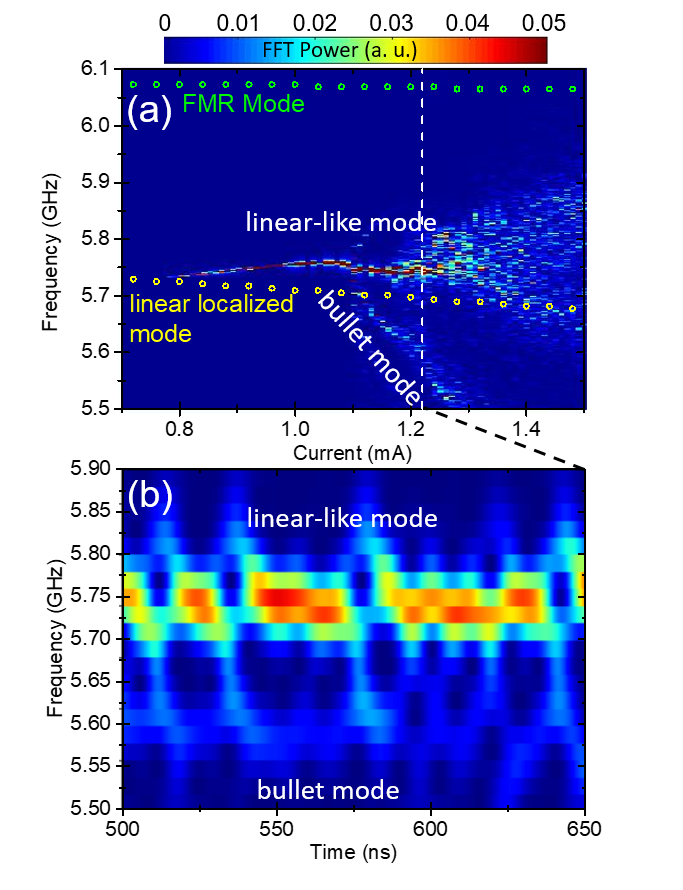}
\caption{\label{fig4} (Color online) (\textit{Simulation}) (a) FFT power colormap of the total magnetization as a function of frequency and current at $H_{\mathrm{IP}}=60$~mT and $\varphi=30\degree$. The eigenmodes, including the FMR and linear localized modes, are shown by green and yellow circles, respectively; (b) spectrogram of total magnetization at $I_{\mathrm{dc}}=1.22$~mA }
\end{figure}

The auto-oscillation spectra vs.~current, and the splitting into two modes can be seen in Fig.~\ref{fig4}~(a), in good agreement with our experimental observations (Fig.~\ref{fig2}). We additionally calculated the linear eigenmodes of the SHNOs by turning off the spin torque in our simulation and exciting the system with a magnetic field pulse (a sinc function with an amplitude of 0.5~mT and a duration of 25~ps). The FMR mode and the linear localized mode are shown in Fig.~\ref{fig4}~(a) by green and yellow circles, respectively. Both auto-oscillation modes are far below the FMR frequency, \emph{i.e.},~both are localized. In agreement with Ref. \onlinecite{Dvornik2018prappl}, the frequency of the auto-oscillation at its onset coincides with the eigenmode of the constriction. In clear contrast, the lower frequency mode that emerges at around 1.1 mA cannot be attributed to the eigenmodes of the constriction, and hence not to any deepening of the spin-wave well caused by the Oersted field.\cite{Demidov2014} Instead, we conclude that the lower frequency mode is a self-localized spin-wave bullet. In contrast to the in-plane magnetized extended films---where bullets typically have lower threshold currents than the quasilinear propagating spin waves---they require higher currents than the linear-like modes of the constriction. This could be attributed to the fact that self-localization in the given volume occurs only when some critical number of magnons is achieved \cite{PhysRevLett.95.237201}, while field confinement happens even for spin waves with vanishing amplitudes. 

It is worth mentioning that the bullet appears to be splitting from the linear-like mode. We, therefore, inspected the transient behaviour of the magnetization dynamics in a multimode regime ($I_{\mathrm{dc}}=1.22$) by performing time-frequency analysis using a short-time Fourier transform. Due to the small frequency gap between the linear-like and bullet modes, we did not decrease the moving window length to less than 50~ns, in order to maintain a reasonable frequency resolution (Kaiser window with $\beta=30$ and overlap of 49.95~ns). As can be seen in Fig.~\ref{fig4}~(b), the discontinuities in the linear-like mode are followed by a sharp transition to the bullet mode---\emph{i.e.}, mode hopping is observed. We, therefore, conclude that the bullet mode does not branch off, but instead nucleates from the linear-like mode towards a lower frequency. To this end, it could be viewed as a self-localization of the field-localized mode of the constriction.

\begin{figure}[t]
    \centering
\includegraphics[trim=0cm 0cm 0cm 0cm, clip=true,width=3.4in]{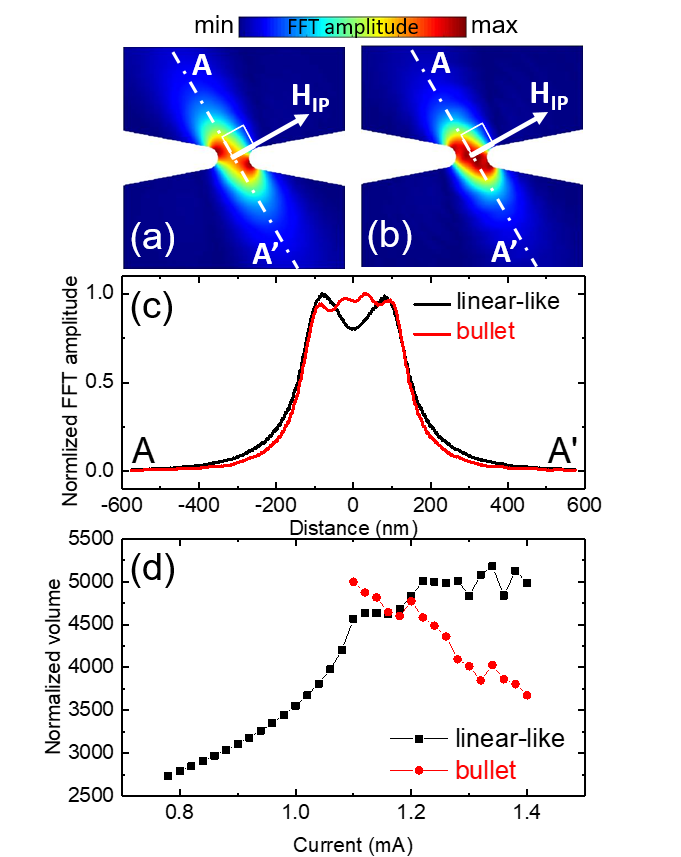}
\caption{\label{fig5} (Color online)
(\textit{Simulation}) The mode profile of the FFT peaks at $H_{\mathrm{IP}}=60$~mT, $\varphi=30\degree$, and $I_{\mathrm{dc}}=1.26$~mA are shown in (a) for the linear-like mode confined to the edges of the nanoconstriction, and (b) for the bullet mode, which is more localized centrally in an area comparable to the constriction size. (c) FFT amplitude along the A--A' direction on the mode profiles perpendicular to the direction of the external field. (d) Normalized volumes of the linear-like mode (black rectangles) and bullet mode (red circles) versus current.}
\end{figure}

To determine the spatial distribution of the observed modes, we performed pointwise temporal FFT over the dynamic component of the magnetization sampled in the vicinity of the constriction---that is, mode profile analysis using the \textsc{semargl-ng} package\cite{dvornik2011numerical, Dvornik2013}. The profiles of the linear-like and bullet modes calculated for the applied current of $I_{\mathrm{dc}}=1.26$~mA are shown in Fig.~\ref{fig5}~(a) and (b), respectively. While the linear-like mode is confined to the edges of the constriction where the internal field has local minima, the bullet is more localized in the center of the constriction, where the internal field instead has a local maximum. This again confirms its predominant self-localization character. 

Similar to what was observed in Refs.\onlinecite{Demidov2014,Kendziorczyk2016prb}, both the linear-like and bullet modes mostly extend along the direction perpendicular to the applied field (shown by the dotted lines in Figs.~\ref{fig5}~(a) and (b)). However, we note that the bullet mode has somewhat lower extent (a smaller ``halo'' is seen in Fig.~\ref{fig5}~(b)). This is confirmed by comparing the cross-sections of the linear-like and bullet modes extracted along the major axes of their profiles (as shown by the dotted lines in Figs.~\ref{fig5}~(a) and (b)). In fact, compared to the linear-like mode, the auto-oscillation power of the bullet mode drops faster with distance from the constriction center. The self-localization thus compresses the bullet's profile. 

To compare the degree of localization of the linear-like and bullet modes, we estimate their auto-oscillation volumes at each current by integrating the corresponding spatial profiles, as explained in Ref.\onlinecite{Dvornik2018prappl}. The calculated volumes are mapped against the bias current in Fig.~\ref{fig5} (d). First, we observe that the volume of the linear-like mode increases, starting from its onset and continuing until the bullet nucleates. Once the multimode state is achieved, the volume of the linear-like mode stays relatively constant, while its slight negative frequency vs.~current slope resembles that of the eigenmode, and so could be attributed to the effect of the Oersted field. In contrast, the volume of the bullet mode drops monotonically with the applied current. According to Ref. \onlinecite{dvornik2018anomalous}, this non-monotonic re-localization of the auto-oscillations is due to the competition of the repulsion and attraction of magnons, caused by the reduction of the static demagnetizing field (shallowing of the spin wave wells) and enhancement of the dynamic dipolar coupling, respectively. It has been predicted that the repulsion of magnons dominates at small amplitudes, consistent with the initially positive nonlinearity of the auto-oscillations observed in Fig.~\ref{fig4}(a). The nucleation of the bullet, thus, indicates that the attraction process becomes dominant. In the vicinity of this point, the auto-oscillations show zero nonlinearity, consistent with the minima in the generational linewidth seen in Fig.~\ref{fig2}(d).\cite{kim2008prl} Although the bullet's frequency drops well below the edge mode, its volume remains considerably higher. This again highlights the difference in the confinement mechanisms of the linear-like and bullet modes, being of a static and dynamic nature, respectively.

\section{\label{sec:Conclusion}Conclusion}

We have determined the dynamic magnetic properties of nanoconstriction-based SHNOs subject to weak in-plane magnetic fields, by carrying out microwave measurements and comparing them with the micromagnetic simulation of the same structure. We found that, at high bias currents, the auto-oscillations spectra show two dominant modes: a linear-like mode that is confined to the edges of the constrictions, and a bullet mode that is confined to the center. While the former is a field-localized mode confined to the minima of the internal field, and therefore slightly depends on the strength of the current and its relative angle with regards to the external field, the frequency of the bullet rapidly decreases with its amplitude. Our simulations reveal that the two modes do not coexist at a given point in time, but instead mode hop, which also manifests itself in the experimentally observed linewidth broadening of the modes. Finally, our simulations have shown that the bullet nucleates from the linear-like mode and then experiences substantial auto-oscillation volume compression due to the nonlinear self-localization. Our findings provide a better understanding of the dynamics of nanoconstriction-based SHNOs necessary for subsequent studies of these systems, including their low-field operation and possible in-plane mutual synchronization.

\section*{Acknowledgments}

This work was supported by the Swedish Foundation for Strategic Research (SSF), the Swedish Research Council (VR), and the Knut and Alice Wallenberg foundation (KAW). This work was also supported by the European Research Council (ERC) under the European Community's Seventh Framework Programme (FP/2007-2013)/ERC Grant 307144 ``MUSTANG''.

\bibliographystyle{aipnum4-1}
\end{document}